\documentclass[twocolumn,prd,showpacs,amsmath,amssymb,floatfix,amsmath,amssymb,superscriptaddress]{revtex4-1}

\usepackage{amssymb}
\usepackage{amsmath}
\usepackage{hyperref}
\usepackage{verbatim}
\usepackage{graphicx}
\usepackage{bm} 
\usepackage{dcolumn}  
\usepackage{epsfig}
\usepackage{color}
\usepackage[usenames,dvipsnames]{xcolor}

\newcommand{\Av}[1]{{\bf #1}}

\newcommand{\Mc}[1]{{\mathcal #1}}

\def\rmd{{\mathrm{d}}}

\def\kB{k_{\mathrm{B}}}

\begin{document}

\title{A molecular dynamics approach to dissipative relativistic hydrodynamics: propagation of fluctuations}

\author{Leila Shahsavar}
\affiliation{Department of Physics, College of Sciences, Shiraz University, Shiraz 71454, Iran}
\author{Malihe Ghodrat}
\email{m.ghodrat@modares.ac.ir}
\affiliation{Department of Physics, College of Sciences, Shiraz University, Shiraz 71454, Iran}
\affiliation{Research Institute for Astronomy and Astrophysics of Maragha (RIAAM), P.O. Box 55134-441, Maragha, Iran}
\affiliation{Department of Physics, Faculty of Basic Sciences, Tarbiat Modares University, P.O. Box 14115-175, Tehran, Iran}

\author{Afshin Montakhab}
\email{montakhab@shirazu.ac.ir}
\affiliation{Department of Physics, College of Sciences, Shiraz University, Shiraz 71454, Iran}

\begin{abstract}
Relativistic generalization of hydrodynamic theory has attracted much attention from a theoretical point of view. However, it has many important practical applications in high energy as well as astrophysical contexts. Despite various attempts to formulate relativistic hydrodynamics, no definitive consensus has been achieved. In this work, we propose to test the predictions of four types of \emph{first-order} hydrodynamic theories for non-perfect fluids in the light of numerically exact molecular dynamics simulations of a fully relativistic particle system in the low density regime. In this regard, we study the propagation of density, velocity and heat fluctuations in a wide range of temperatures using extensive simulations and compare them to the corresponding analytic expressions we obtain for each of the proposed theories. As expected in the low temperature classical regime all theories give the same results consistent with the numerics. In the high temperature extremely relativistic regime, not all considered theories are distinguishable from one another. However, in the intermediate regime, a meaningful distinction exists in the predictions of various theories considered here. We find that the predictions of the recent formulation due to Tsumura-Kunihiro-Ohnishi are more consistent with our numerical results than the traditional theories due to Meixner, modified Eckart and modified Marle-Stewart.
\end{abstract}



\maketitle

\section{Introduction}
The special relativistic generalization of Newtonian hydrodynamics has attracted much attention in both statistical physics and high energy physics since its early days \cite{Jut11,Eck40,Isr63,Muel67,Stew71,Isr79,Lan87} and has found applications in a wide range of physical processes from astrophysical phenomena \cite{Rez13,Lop11,Bal01}, to the hydrodynamic description of the high temperature quark-gluon (QG) plasma in the heavy-ion collision experiments at CERN and BNL \cite{Huo04,Kol04}, to the recent studies on graphene \cite{Mul08a,Mul09,Men11,Tor15}. Indeed, the growing interest in relativistic hydrodynamic (RH) is not restricted to the hydrodynamics of perfect fluids \cite{Lan87,Cer02} but also extends to dissipative description of relativistic systems \cite{Rez13,Mur02,Rom10,Jai13,Jai13b,Jai13c,Jai14}

Despite its long history and wide usage, there are still disagreements on the fundamental postulates and definitions of the theory, specifically in the presence of dissipative effects. One key source of the ambiguity is the definition of the basic variables of the theory such as the hydrodynamic velocity four-vector, the dissipative part of energy-momentum tensor, as well as the constraints that should be placed on them. Another source of disagreements is the derivatives expansion of the entropy current in terms of hydrodynamic variables. In what is referred to as \emph{first-order} theories, entropy current contains terms that are first-order in terms of thermodynamic fluxes \cite{Eck40,Lan87,Rez13}. The responses to deviations from equilibrium states in these theories are considered to be linearly proportional to thermodynamic forces and hence are instantaneous in nature (i.e. no relaxation times are assumed). These features give rise to unphysical instabilities \cite{Lind85} especially in Eckart frame and/or \emph{parabolic} set of hydrodynamic equations which lead to causality problems \cite{Rez13}. It should be noted that while the former problem seems to be due to the choice of frame, the latter issue is a fundamental pitfall which is also present in classical hydrodynamics. Nevertheless, both issues were historically the main motivations for adopting non-relativistic \cite{Muel67, Liu83} and relativistic \cite{Isr76,Stew77,Liu86} extended second-order theories \cite{Rez13}.

In addition to extended theories, attempts have also been made in the context of first-order theories \cite{Gcol06,Gper09b,Tsum07,Bir12} to address or resolve the first two drawbacks associated with RH; namely the ambiguities in definition of fundamental variables and the unstable features that arise in the Eckart frame. The relativistic extension  of Meixner's idea (MX) \cite{Gcol06,Meix41,Meix43}, the modified Eckart theory (ME) \cite{Cer02,Gper09b}, the modified version of Marle-Stewart original proposal (MMS) \cite{Stew71} and the recent rigorous approach on the basis of renormalization group (RG) method presented by Tsumura-Kunihiro-Ohnishi (TKO) \cite{Tsum07} are examples of these efforts that we intend to investigate in this work.

Despite the difficulties associated with the general class of the first-order theories, it is important to realize that such theories provide a minimal theoretical model which are believed to provide fairly accurate description of relativistic fluids in the long-wavelength hydrodynamic limit (where causality becomes less relevant) and a static background (where the above-mentioned theories are known to be stable \cite{Pu10}). However, we note that there exist other first-order theories \cite{Bir12} which are stable in a generic frame and will not be discussed in the present work.

The precision of various second-order theories has been tested in the context of numerical solution of the relativistic Boltzmann equation \cite{Huo09,Bou09,Den10,Den14}. However, here, we propose to study the accuracy of the above-mentioned first-order theories in the light of the microscopic approach of molecular dynamics of a hard-sphere gas, which to the best of our knowledge has not been done before. Also, since each of these theories are based on a different set of assumptions, a test of their accuracy can also provide a prospective for their basic assumptions. To achieve our goal, we focus on propagation of fluctuations and obtain analytical expressions which can then be used to calculate how such fluctuations propagate according to various theories. Such predictions are then compared to the predictions of numerically exact results obtained from our model in a wide range of temperatures. Our results indicate that MX is the least accurate theory, and while the predictions of ME and MMS are fairly accurate (and the same), it is the recently proposed TKO theory that provides the best fit to the data for a wide range of temperature regime.

This paper is organized as follows: Section II provides a theoretical background to our work presenting the essential concepts of hydrodynamic theory and its relativistic generalizations to special relativity. In Section III, the details of our numerical experiments are presented. The analytical predictions are compared to simulation data in Section IV and we shall conclude in Section V.

\section{Theoretical Background}
\subsection{Hydrodynamic equations of a relativistic fluid}\label{s:Hyd_discript}
Hydrodynamic description of a fluid is appropriate when observation time and length scales are much larger than the mean free time (MFT) and mean free path. In such a case the system is described in terms of a few functions, defined at each point in space and time. For a single-component fluid in low energy regimes, for example, these functions correspond to the mass density $\rho(\Av r,t)$, the velocity field $\Av v(\Av r,t)$, and two thermodynamic variables, e.g., pressure $\Mc P(\Av r,t)$ and temperature $T(\Av r,t)$. In the extension of the classical hydrodynamics to special relativity, the fluid velocity, $\Av v$, is replaced by the Lorentz covariant four-vector, $u^\mu:=dx^\mu/d\tau$, which satisfies the normalization condition, $u_\mu u^\mu=-1$ with $\tau$ denoting the fluid proper time.

In response to small perturbations, the evolution of the fluid towards its equilibrium state is governed by the so called \emph{hydrodynamic equations} which are in fact balance equations for conserved quantities including mass, momentum and energy densities.
For the special case of a \emph{perfect fluid}, one may define particle current, $\Mc J^\mu=\rho u^\mu$, and energy-momentum tensor, $\Mc E^{\mu\nu}= (e+\Mc P)u^\mu u^\nu+\Mc P g^{\mu\nu}$, obeying the conservation laws, $\partial_\mu \Mc J^\mu =0$ and $\partial_\mu \Mc E^{\mu\nu}=0$, in analogy with classical hydrodynamics. The basic equations of relativistic perfect fluid are then given by
\begin{eqnarray}\label{e:Hyd_eqs_rel_pf}
\partial_\mu (\rho u^\mu) &=&0 \qquad \textrm{Continuity} \label{e:Hyd_eqs_rel_pf_a}\\
\rho h a_\nu + \Delta^\mu_{\nu} \partial_\mu \Mc P&=&0 \qquad \textrm{Euler}
\label{e:Hyd_eqs_rel_pf_b}\\
 u^\mu\partial_\mu e + \rho h \Theta &=&0 \qquad \textrm{Energy balance}
 \label{e:Hyd_eqs_rel_pf_c}
\end{eqnarray}
in which $e$ and  $\rho h= e+\Mc P$ are respectively, total energy and enthalpy densities, $a_\nu:=u^\mu\partial_\mu u_\nu$ denotes the hydrodynamic acceleration, $\Theta:=\partial_\mu u^\mu$ is the relativistic counterpart of the classical fluid expansion scalar, $\Delta_{\mu\nu}=g_{\mu\nu}+u_\mu u_\nu$, defines the projection operator in the direction perpendicular to the hydrodynamic velocity, and $g_{\mu\nu}=\mathrm{diag}(-1,1,1,1)$ represents the metric tensor \cite{Rez13}. Together with the equation of state of the fluid, the above equations give a complete description of a relativistic perfect fluid.

The extension to the hydrodynamic equations of a non-perfect (NP) fluid is achieved by adding the contribution of viscosity and thermal-conductivity effects to the fundamental variables previously defined for a perfect (P) fluid, i.e., $\Mc J^\mu=\Mc J^\mu_{P}+\Mc J^\mu_{NP}$
and $\Mc E^{\mu\nu}=\Mc E^{\mu\nu}_{P}+ \Mc E^{\mu\nu}_{NP}$. Derivation of these terms in a relativistically consistent manner, however, is far from a straightforward procedure. The reason could be traced back to the ambiguities associated with the fundamental definitions as well as the postulates of the theory. For instance, a unique bulk velocity, and thus a well-defined frame comoving with the fluid cannot be introduced in the presence of viscosity and heat conduction, since in this case the local rest frame velocity does not coincide with the mean macroscopic velocity. Therefore, it is traditionally admitted to base the theory on either Eckart or Landau definition of hydrodynamic velocity. The former, also known as particle frame, assumes that velocity four-vector is parallel to the particle current and thus the dissipative parts of particle current and energy-momentum tensor satisfy the conditions $u_\mu u_\nu \Mc E^{\mu\nu}_{NP}=0$, $u_\mu \Mc J^\mu_{NP}=0$ and $\Delta_{\mu\nu}\Mc J^\nu_{NP}=0$ \cite{Gro80,Kam87}. In the Landau energy frame, the first two constraint hold without change while the third one is replaced by, $u_\mu \Delta_{\nu\lambda}\Mc E^{\mu\nu}_{NP}=0$, indicating the four-velocity is the eigenvector of energy-momentum tensor. Note that these two definitions of velocity four-vector are related to one another through a heat current vector in general and they are equal in the perfect fluid picture where non-diagonal terms are absent in energy-momentum tensor \cite{Rez13,Cer02}. To complete our list, it is worthwhile to mention a different set of constraints, proposed by Marle-Stewart \cite{Stew71}, in Eckart particle frame, ${\Mc E_{NP}}^{\mu}_{\,\mu}=0$, $u_\mu \Mc J^\mu_{NP}=0$, $\Delta_{\mu\nu}\Mc J^\nu_{NP}=0$, which requires the energy-momentum tensor to be traceless.

A careful study on the microscopic origins and physical meaning of these assumptions calls for fundamental approaches to derive the phenomenological hydrodynamic equations from the microscopic equations as is done in derivation of relativistic Boltzmann equation \cite{Isr63,Stew71,Isr79,Gro80} using Chapman-Enskog method \cite{Chap39} or Grad's fourteen-moment method \cite{Gra49}. In a recent systematic approach on the basis of RG method \cite{Tsum07}, the authors introduce a wider class of frames in which particle frame (Eckart) and energy frame (Landau) are regarded as special cases. Their analysis reveals that the resulting equations in energy frame is consistent with Landau-Lifshitz constraints, while in the particle frame, the constraints by Marle-Stewart as opposed to Eckart must be used. However, the resulting hydrodynamical equation which manifestly satisfies the second law of thermodynamics is neither similar to that of Eckart nor to that of Marle-Stewart. We next briefly discuss various possible hydrodynamic formalisms in the Eckart particle frame and compare the resulting set of hydrodynamic equations obtained.

In the Eckart formalism, the particle current is assumed to be parallel to the velocity four-vector, i.e., $\Mc J^{\mu}=\rho u^\mu$, and the energy-momentum tensor is defined as
\begin{equation}\label{e:Eck_EMtensor}
\Mc E^\mu_\nu=e u^\mu u_\nu+(\Mc P+\Pi)\Delta^\mu_\nu+\pi^\mu_\nu+q^\mu u_\nu+u^\mu q_\nu,
\end{equation}
where $\Pi$, $\pi^\mu_\nu$ and $q^\mu$ are thermodynamic fluxes, respectively known as viscous bulk pressure, anisotropic shear tensor, and heat flux. The entropy flux is also generalized to $\Mc S^\mu= \rho s u^\mu + \frac{1}{T}\Mc R^\mu$, with $\Mc R^\mu$, accounting for dissipative effects. It turns out that $\Mc R^\mu=q^\mu$ in first-order theories while it might include higher order terms in the extended theories \cite{Rez13}. The condition of positive entropy production, $\partial_\mu \Mc S^\mu \geq0$, together with the continuity equations obtained from conservation laws, would then lead to the \emph{constitutive equations} \cite{Rez13}:
\begin{eqnarray}
\Pi&=&-\zeta\Theta \label{e:constv_eck_a}\\
\pi^\mu_\nu&=&-2\eta\sigma^\mu_\nu \label{e:constv_eck_b}\\
q^\mu&=&-\lambda T(\Mc D^\mu \ln T + a^\mu), \label{e:constv_eck_c}
\end{eqnarray}
where $\sigma^\mu_\nu=\frac{1}{2}(\Mc D^\mu u_{\nu}+\Mc D_{\nu} u^{\mu})- \frac{1}{3}\Theta \Delta^{\mu}_{\nu} $ is the relativistic shear tensor and $\Mc D^\mu=\Delta^{\nu\mu}\partial_\nu$. The new parameters, $\zeta$, $\eta$ and $\lambda$, respectively denote bulk viscosity, shear viscosity and thermal conductivity whose explicit expressions are derived from an appropriate kinetic theory such as Boltzmann theory. This would thereby close the set of hydrodynamic equations for a non-perfect fluid.

The appearance of the fluid acceleration as a driving force in the expression of heat flux leads to nonzero heat flux in the absence of temperature gradient. It is this purely relativistic effect in Eckart's theory which gives rise to exponential growth of small fluctuations \cite{Lind85}. In the following, we shall briefly discuss some of the attempts that have been made to alleviate these unphysical instabilities in the context of first-order theories in a \emph{static background}.

The first widely used proposal to make a stable theory out of the pioneering work of Eckart is to approximate the acceleration term by a pressure gradient using the Euler equation (Eq.~\eqref{e:Hyd_eqs_rel_pf_b}). This would lead to what is called ``modified Eckart'' formulation \cite{Gper09b,Cer02} with the following expression for the heat flux,
\begin{equation}
q^\mu=-\lambda T(\Mc D^\mu \ln T - \frac{1}{\rho h} \Mc D^\mu \Mc P),
\end{equation}
which considering the equation of state of an ideal fluid, $\Mc P= \rho T$, is representable in the form of Fourier law, $\mathbf{q}=-\kappa \nabla T$, and thus revealing the parabolic nature of the theory \cite{Gho13}. However, it leads to a set of hydrodynamic equations with stable equilibrium in a static background (see next section for more details).

The second approach is referred to as Meixner's formalism \cite{Gcol06} and differs from the original Eckart theory in its definition of energy-momentum tensor which reads as
\begin{equation}\label{e:Meix_EMtensor}
\Mc E^\mu_\nu=e u^\mu u_\nu+(\Mc P+\Pi) \Delta^\mu_\nu+\pi^\mu_\nu.
\end{equation}
This formalism is actually a relativistic generalization of Newtonian hydrodynamic equations \cite{Meix41} in which the heat energy is not included in the Navier-Stokes equation but appears in the energy balance equation. The heat flux is therefore derived solely by a temperature gradient, $q^\mu=-\lambda T \Mc D^\mu \ln T$,  as is the case in Newtonian hydrodynamics. The other two constitutive equations are the same as Eqs.~\eqref{e:constv_eck_a}, \eqref{e:constv_eck_b} leading to stable set of equations \cite{Gper09b}.

The third formalism is Marle-Stewart proposal which requires to change the original constraint on the energy-momentum tensor used by Eckart to ${\Mc E_{NP}}^{\mu}_{\,\mu}=0$. Consequently the new energy momentum tensor in this formalism is given by
\begin{equation}\label{e:ene_MS}
    \Mc E_{\nu}^{\mu}=(e+3\Pi) u^{\mu} u_{\nu}+(\Mc P+\Pi) \Delta^\mu_\nu+\pi^\mu_\nu+q^{\mu} u_{\nu}+u^{\mu} q_{\nu},
\end{equation}
with the bulk viscous pressure given by $\Pi=+\zeta (3\gamma-4)^{-1}\Theta$, in which $\gamma=c_\Mc P/c_{\small V}$ and $c_{\small V}$ ($c_\Mc P$) denotes the heat capacity at constant volume (pressure). The anistropic shear tensor and heat flux are the same as given in Eqs. \eqref{e:constv_eck_b} and \eqref{e:constv_eck_c}. In analogy with modified Eckart formalism, one may resolve the instabilities caused by the acceleration term using Euler equation, and in this sense, the resulting stable theory should be named as ``modified'' Marle-Stewart theory.

The final approach is proposed by Tsumura et~al.~\cite{Tsum07}, where the set of constraints on the current and energy-momentum are shown to be the same as that of Marle-Stewart while the constitutive equations to be substituted in Eq.~\eqref{e:ene_MS} are given as, $\Pi=-\zeta (3\gamma-4)^{-2}\Theta$, $
\pi^\mu_\nu=-2\eta\sigma^\mu_\nu$ and $q^{\mu}=-\lambda T \Mc D^\mu \ln T$. Note that although the heat energy is included in the energy-momentum tensor, the acceleration term is absent in the heat flux expression giving rise to a stable set of hydrodynamic equations \cite{Tsum08}.

\begin{table*}[t!]
\begin{center}
\begin{tabular}{|l|ll|ll|}
\hline
&&&&\\
Theory && Basic equations && Sound attenuation parameter \\
\hline
&&&&\\
 ME && $\Mc E^\mu_\nu=e u^\mu u_\nu+(\Mc P+\Pi)\Delta^\mu_\nu+\pi^\mu_\nu+q^\mu u_\nu+u^\mu q_\nu$ &&$\Gamma=\frac{1}{2}[b_0+\frac{L_{T}}{\rho T c_{V}}(1-\frac{1}{\gamma}-\frac{\beta T}{\kappa_{T}\rho h})+\frac{L_{\rho}}{\rho}(\frac{\beta}{c_\Mc P }-\frac{1}{ h })]$\\

 && $q^\mu=-\lambda T(\Mc D^\mu \ln T - \frac{1}{\rho h} \Mc D^\mu \Mc P)$,\;$\Pi = -\zeta_0 \Theta$,\; $\pi^\mu_\nu=-2\eta\sigma^\mu_\nu$ && \\

  &&&& \\
   MX && $\Mc E^\mu_\nu=e u^\mu u_\nu+(\Mc P+\Pi) \Delta^\mu_\nu+\pi^\mu_\nu$ &&$\Gamma=\frac{1}{2}[b_0+\chi(\gamma-1)]$ \\

   && $q^\mu = -\lambda T \Mc D^\mu \ln T$,\;$\Pi = -\zeta_0 \Theta$ ,\; $\pi^\mu_\nu=-2\eta\sigma^\mu_\nu$ && \\

  &&&&\\
  MMS && $\Mc E^\mu_\nu=(e+3\Pi) u^{\mu} u_{\nu}+(\Mc P+\Pi) \Delta^\mu_\nu+\pi^\mu_\nu+q^{\mu} u_{\nu}+u^{\mu} q_{\nu}$ &&$\Gamma=\frac{1}{2}[b_1+\frac{L_{T}}{\rho T c_{V}}(1-\frac{1}{\gamma}-\frac{\beta T}{\kappa_{T}\rho h})+\frac{L_{\rho}}{\rho}(\frac{\beta}{c_\Mc P}-\frac{1}{h})-\frac{3 \beta \zeta_1}{\rho^{2} c_{V} \kappa_{T} h}]$ \\

   && $ q^\mu=-\lambda T(\Mc D^\mu \ln T - \frac{1}{\rho h} \Mc D^\mu \Mc P)$,\;$\Pi=-\zeta_1 \Theta$ ,\; $\pi^\mu_\nu=-2\eta\sigma^\mu_\nu$ && \\

  &&&& \\
  TKO && $\Mc E^\mu_\nu=(e+3\Pi) u^{\mu} u_{\nu}+(\Mc P+\Pi) \Delta^\mu_\nu+\pi^\mu_\nu+q^{\mu} u_{\nu}+u^{\mu} q_{\nu}$ &&$\Gamma=\frac{1}{2}[b_2+\frac{\lambda}{\rho c_{V}}(1-\frac{1}{\gamma}-\frac{\beta T}{\kappa_{T}\rho h})-\frac{3 \beta \zeta_2}{\rho^{2} c_{V} \kappa_{T} h}]$\\

   && $q^\mu = -\lambda T \Mc D^\mu \ln T$,\;$\Pi=-\zeta_2 \Theta$,\; $\pi^\mu_\nu=-2\eta\sigma^\mu_\nu$ &&\\
   &&&&\\
 \hline
\end{tabular}
\caption{Fundamental equations and relevant parameters for different first-order theories. The thermal diffusivity, $\chi=\lambda/(\rho c_\Mc P)$, is identical in all formalisms. The transport coefficients, $\lambda, \eta, \zeta$ for hard-sphere gas are given in \cite{Cer02} and the equation of state we have used is $\Mc P (\upsilon-b)= T$ with $ b=2\pi d^{3}/3$ and $ \upsilon=V/N $. Thermal expansion coefficient and isothermal compressibility are respectively given by $\beta=\kappa_T \beta_V $, and $ \kappa_T=-\frac{1}{V}(\partial V/\partial \Mc P)_T$. Other coefficients are $b_i=(3\zeta_i+4\eta)/3\rho h$ (with $i=0,1,2$), $\zeta_0=\zeta$, $\zeta_1=-\zeta/(3\gamma-4)$, $\zeta_2=\zeta/(3\gamma-4)^2$, $L_{T}=\lambda T (1+\beta T/\rho h \kappa_{T})$, $L_{\rho}= \lambda T/(\rho h \kappa_{T})$, $\gamma=c_\Mc P/c_V$, $c_\Mc P=c_V+T$.}
\label{table_1}
\end{center}
\end{table*}

\subsection{Propagation of fluctuations}\label{s:fluc_propag}
The above mentioned hydrodynamic theories give different mechanisms for propagation of perturbations in the fluid. We intend to provide a realistic numerical laboratory to test the predictions of such theories in this regard. In this section we describe how such perturbations are formulated. To achieve this, we use a frame comoving with the fluid in which the mean hydrodynamic four-velocity is $\bar{u}^\mu=u^\mu_0=(1,\mathbf{0})$ and the basic state variables are chosen to be $\rho$, $T$ and $u^\alpha$ with $\alpha=1,2,3$. Introducing small perturbations about the equilibrium state $\Mc A=\Mc A_0+ \delta \Mc A$ (i.e. $\rho=\rho_0+\delta\rho, T=T_0+\delta T, u^\alpha=\delta u^\alpha$) in the hydrodynamic equations and keeping the linear terms with respect to deviations \cite{Cer02,Rez13} would lead to dynamical equations for the evolution of fluctuations. After a standard Fourier-Laplace transform with respect to space and time,
\begin{equation}
     \hat{\Mc A}_k(\omega)=\int_0^\infty \exp(-\omega t) dt \int \delta \mathcal{A} (\Av r,t) \exp(i\Av k\cdot \Av r)d\Av r,
\end{equation}
and decomposing hydrodynamic velocity to longitudinal (parallel to $\mathbf k$) and transverse (perpendicular to $\mathbf k$) components, the set of hydrodynamic equations can be cast in the form, $\hat{\Mc O}_k (\omega)=\mathbf{M}^{-1}(k,\omega)\Mc O_k$, relating the state vector $\hat{\Mc O}_k(\omega):= [\hat{\rho}_{k}(\omega),\hat{T}_{k}(\omega),\hat{u}_{k}^{\parallel}(\omega),\hat{\Av u}_{k}^{\perp}(\omega)]$, to its initial value $\Mc O_k$, through
the hydrodynamic matrix $\mathbf{M}$. The dispersion relations for the hydrodynamic collective modes are determined by the complex roots of the equation $\det \mathbf{M}=0$, which in the most general case would give two identical transverse modes, $\omega_t=-\eta k^2/\rho h$, and three longitudinal modes, namely one thermal mode, $\omega_0=- \chi k^2$, and two sound modes, $\omega_{\pm}=\pm ic_{s} k -\Gamma k^2,$ all damping out by viscous and thermal dissipative processes characterized by thermal diffusivity, $\chi$, and sound attenuation parameter, $\Gamma$. This becomes evident by considering the dynamical equations for the evolution of longitudinal modes \cite{Rei80},
\begin{eqnarray}
    \label{e:corr_rho}\frac{\rho_k(t)}{\rho_k(0)}&=&\frac{\gamma-1}{\gamma}e^{-\chi k^2 t} +\frac{1}{\gamma}e^{-\Gamma k^2 t}\cos(c_s k t)\\
    \label{e:corr_u}
    \frac{ u_k^{\parallel}(t)}{u_k^{\parallel}(0)}&=&e^{-\Gamma k^2 t}\cos(c_s k t)\\
    \label{e:corr_Q}\frac{ Q_k(t)}{Q_k(0)}&=&e^{-\chi k^2 t},
\end{eqnarray}
where we have kept terms up to zeroth order of $k$ in amplitude which is an appropriate assumption in the hydrodynamic limit (See the Appendix for first order corrections). Equation \eqref{e:corr_Q} has been obtained by defining heat energy density as $Q(\Av r,t):=e(\Av r,t)-h \rho(\Av r,t),$ \cite{Rez13,Han13} and using the thermodynamic relation, $\delta Q=-\frac{T \beta_{\small V}}{\rho} \delta\rho(\Av r,t)+\rho c_{\small V} \delta T(\Av r,t)$, to write down `heat energy density' fluctuations as a function of temperature and density fluctuations with $\beta_V=(\partial \Mc P/\partial T)_\rho$ being the thermal pressure coefficient.

Eq.~\eqref{e:corr_u} consists of two acoustic waves moving with sound speed $c_s$ in the opposite directions from the center. The sound attenuation parameter, $\Gamma$, controls the width of the sound peaks whose functional form depends on the underlying hydrodynamic equations (see Table 1). Eq.~\eqref{e:corr_Q}, on the other hand, demonstrates the purely diffusive propagation of heat fluctuations throughout the fluid. It is a single peaked function centered at zero whose width is proportional to thermal diffusivity parameter, $\chi=\lambda/(\rho c_{\Mc P} )$, which is identical in all formalisms discussed in Section \ref{s:Hyd_discript}. The propagation of perturbations in density as inferred from Eq.~\eqref{e:corr_rho}, occurs through a combination of diffusive (thermal) and wave-like (acoustic) modes that are respectively referred to as Rayleigh and Brillouin peaks in density correlation spectrum. Table~\ref{table_1} shows the fundamental equations and relevant parameters in each of the theories considered in the present work.

\section{Model and Method}\label{s:simulation}
Our model system \cite{Mon09} consists of $N$ identical impenetrable spherical particles of diameter $d$ and mass $m$ confined in a box of volume $V=L_x L_y L_z$, with periodic boundary conditions. The particles move freely in space until they contact at distance $d$ where they experience a purely repulsive binary interaction,
\begin{equation}
U(r)=\left\{
\begin{array}{cc}
  +\infty, & \; r\leq d \\
  0, & \; r> d \\
\end{array}
\right.
\end{equation}
This hard-core potential model mimics the strong repulsion between the atoms and molecules at small distances and is appropriate to study many features of fluids in and out of equilibrium \cite{All86}. Recently, employing relativistic particle dynamics, it has been shown that the model is an ideal one to simulate and investigate the thermostatistical properties of a relativistic gas due to the unique characteristics of hard-sphere interaction \cite{Gho13,Mon09,Gho11,Gho09}. First, it is a contact potential that overcomes the difficulties associated with interacting relativistic particle \cite{Whe49} and therefore lets us have a fully relativistic model. Second, it is specifically a good model to simulate hadronic particles which are shown to have constant cross-section in a wide range of energies \cite{Gro80}.

In order to obtain the spatiotemporal correlation function (or correlation profile), we coarse-grain the space in $x$ direction by dividing the system into $N_s$ equal slabs of size $L_y L_z \Delta x$ and measure the fluctuation of a thermodynamic parameter, $\Delta \Mc A$, in each slab using the following definition
\begin{equation}
\Delta \Mc A(x_i,t)= \int_{x_i-\frac{\Delta x}{2}}^{x_i+\frac{\Delta x}{2}}\Mc A(x,t)\rmd x-\,\bar{\Mc A},
\end{equation}
where $x_i\in[-\frac{L_x}{2},+\frac{L_x}{2}]$ is the midpoint of the $i$th slab and $\bar{\Mc A}$ denotes the global average of the quantity $\Mc A(x,t)$. The normalized correlation between the fluctuation in the reference slab (which is the middle slab in our simulations) and the effect it induces at another position and at a later time, is defined as
\begin{equation}\label{e:correlation}
C_{\Mc A}(x_i,t)=\frac{\langle\Delta \Mc A(x_i,t)\Delta \Mc A(0,0)\rangle}
{\langle\Delta \Mc A(0,0)\Delta \Mc A(0,0)\rangle}-C_{inh},
\end{equation}
in which $\langle...\rangle$ represents the averaging over equilibrium distribution of fluctuations. The constant, $C_{inh}=1/(1-N_s)$, is the inherent correlation generated in micro-canonical ensembles due to the fact $\Mc A$ is a conserved quantity and $\sum_i \Delta\Mc A(x_i,0)=0$, which is different from the causal correlation we are interested in and thus should be subtracted \cite{zha06}. We note that, the simulation method used is the well-known event driven method \cite{Ald59} with the cell linked-list tool \cite{All86} implemented to reduce the computational time. The spatial and temporal distances are respectively rescaled with the diameter of spherical particles, $d$, and the shortest time scale in the system, i.e. $t^*=d/c$, while we choose to set $m=d=c=\kB=1$ for convenience.

\section{Results}\label{s:Result}
We have performed extensive numerical simulations studying the propagation of fluctuation (via Eq.~\eqref{e:correlation}) for different thermodynamic variables and various temperature regimes, as a function of time. The obtained results are then compared to the back Fourier transform of analytical correlation function, $C_{\mathcal A}(k,t)=\langle{\mathcal A}_k(t)\mathcal A_{-k}(0)\rangle/\langle{\mathcal A}_k(0)\mathcal A_{-k}(0)\rangle$ \cite{Rei80}. Assuming that $\mathbf k$ is in the $x$ direction one obtains the corresponding correlation functions as bellow:

\begin{eqnarray} \label{e:corr_rho_BFT}
  \label{e:corr_rho_BFT} \nonumber C_{\rho}(x,t)&=&\frac{1}{2\sqrt{\pi}}[\frac{\gamma -1}{\gamma}\frac{1}{\sqrt{\chi t}}e^{-\frac{x^{2}}{4 \chi t}}+\frac{1}{ 2\gamma}\frac{1}{\sqrt{\Gamma t}}\times\\
 &&\qquad\quad(e^{-\frac{(x-c_{s} t)^{2}}{4 \Gamma t}}+e^{-\frac{(x+c_{s} t)^{2}}{4\Gamma t}})]\\
 \label{e:corr_u_BFT} \nonumber C_{u}(x,t)&=&\frac{1}{4\sqrt{\pi}\sqrt{\Gamma t}}(e^{-\frac{(x-c_{s} t)^{2}}{4 \Gamma t}}+e^{-\frac{(x+c_{s} t)^{2}}{4\Gamma t}})\\
 \\
 \label{e:corr_q_BFT} C_{Q}(x,t)&=&\frac{1}{2\sqrt{\pi}\sqrt{\chi t}}e^{-\frac{x^{2}}{4 \chi t}}.
\end{eqnarray}

In the following we shall first present a detailed discussion of our results in the low temperature regime and then proceed to the intermediate and relativistic regimes in order to test the accuracy of different relativistic formalisms.

\subsubsection{Low temperature Newtonian regime}
\begin{figure*}[t!]\begin{center}
	\begin{minipage}[h]{0.32\textwidth}\begin{center}
		\includegraphics[width=\textwidth]{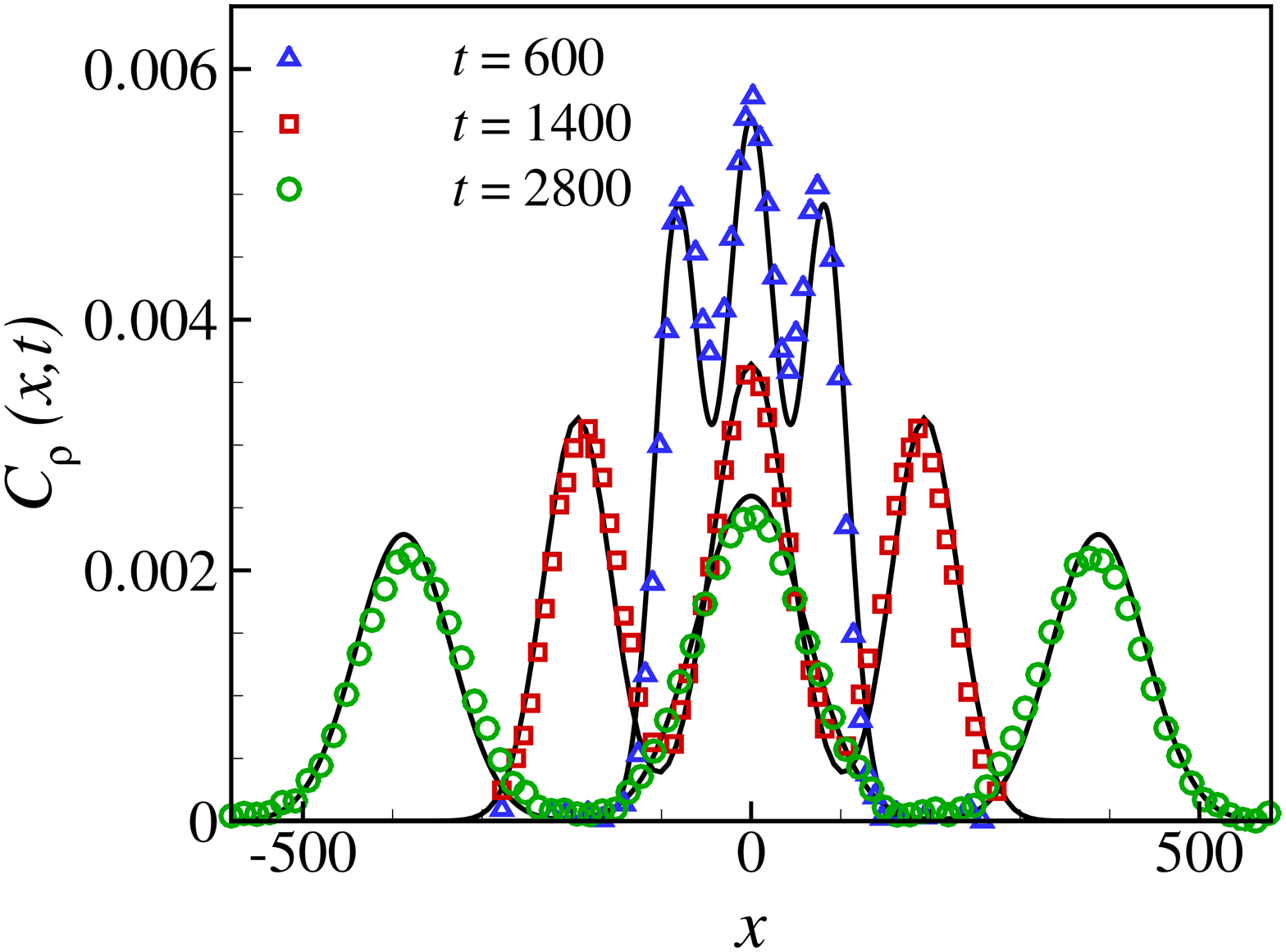} (a)
	\end{center}\end{minipage} \hskip0.01cm	
	\begin{minipage}[h]{0.32\textwidth}\begin{center}
		\includegraphics[width=\textwidth]{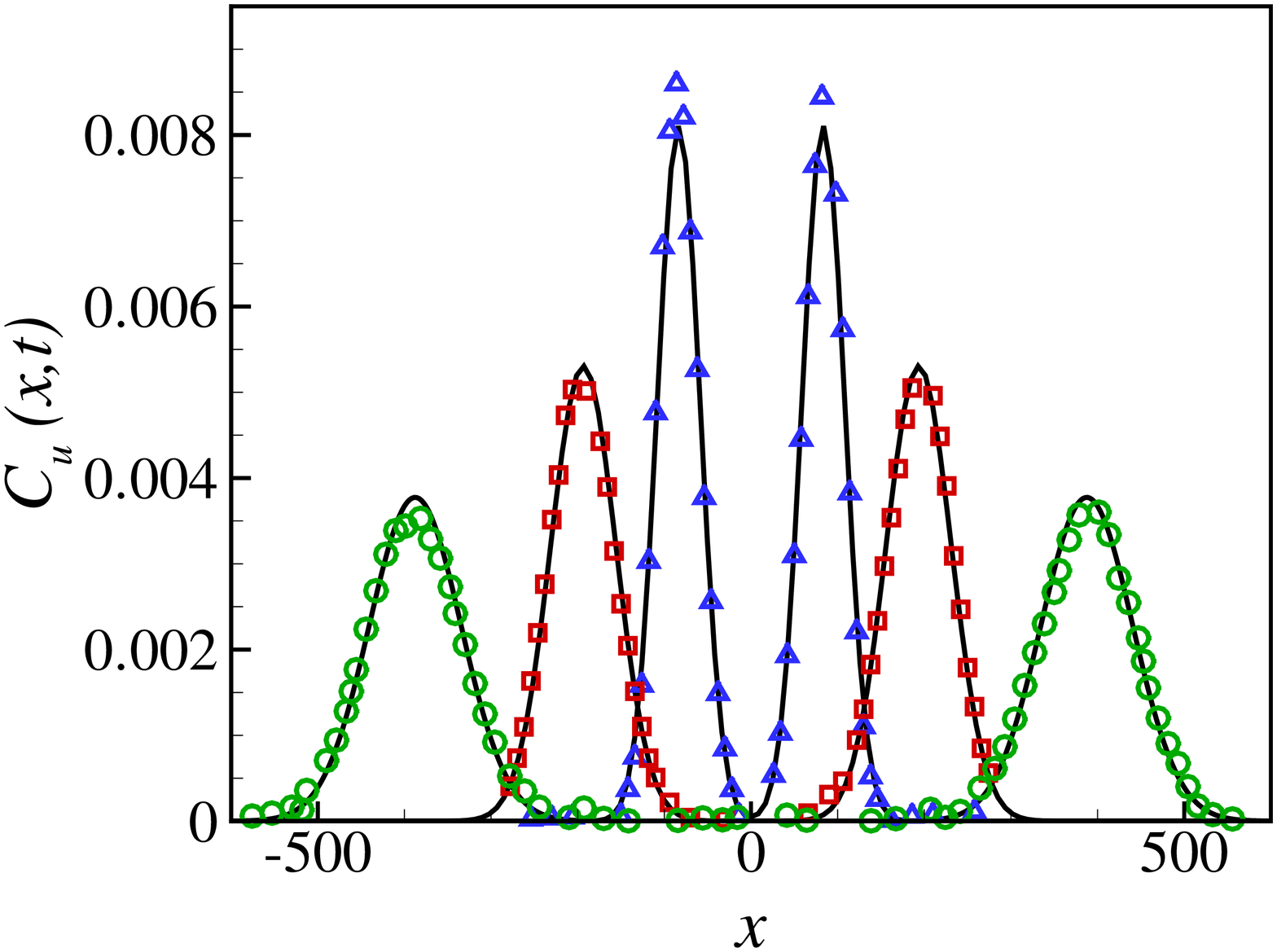} (b)
	\end{center}\end{minipage} \hskip0.01cm	
    \begin{minipage}[h]{0.32\textwidth}\begin{center}
		\includegraphics[width=\textwidth]{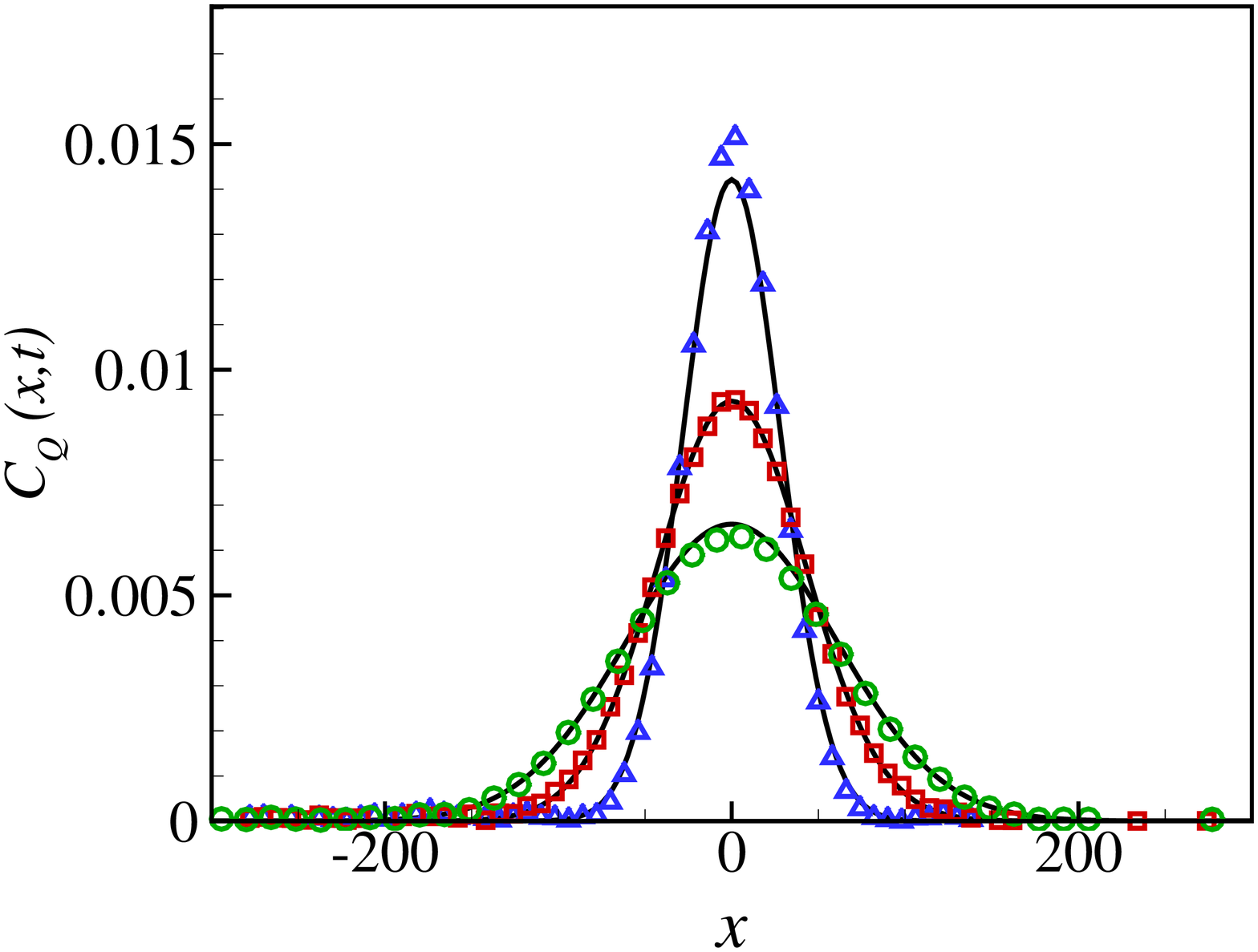}  (c)
	\end{center}\end{minipage} 	\\
\begin{minipage}[h]{0.32\textwidth}\begin{center}
		\includegraphics[width=\textwidth]{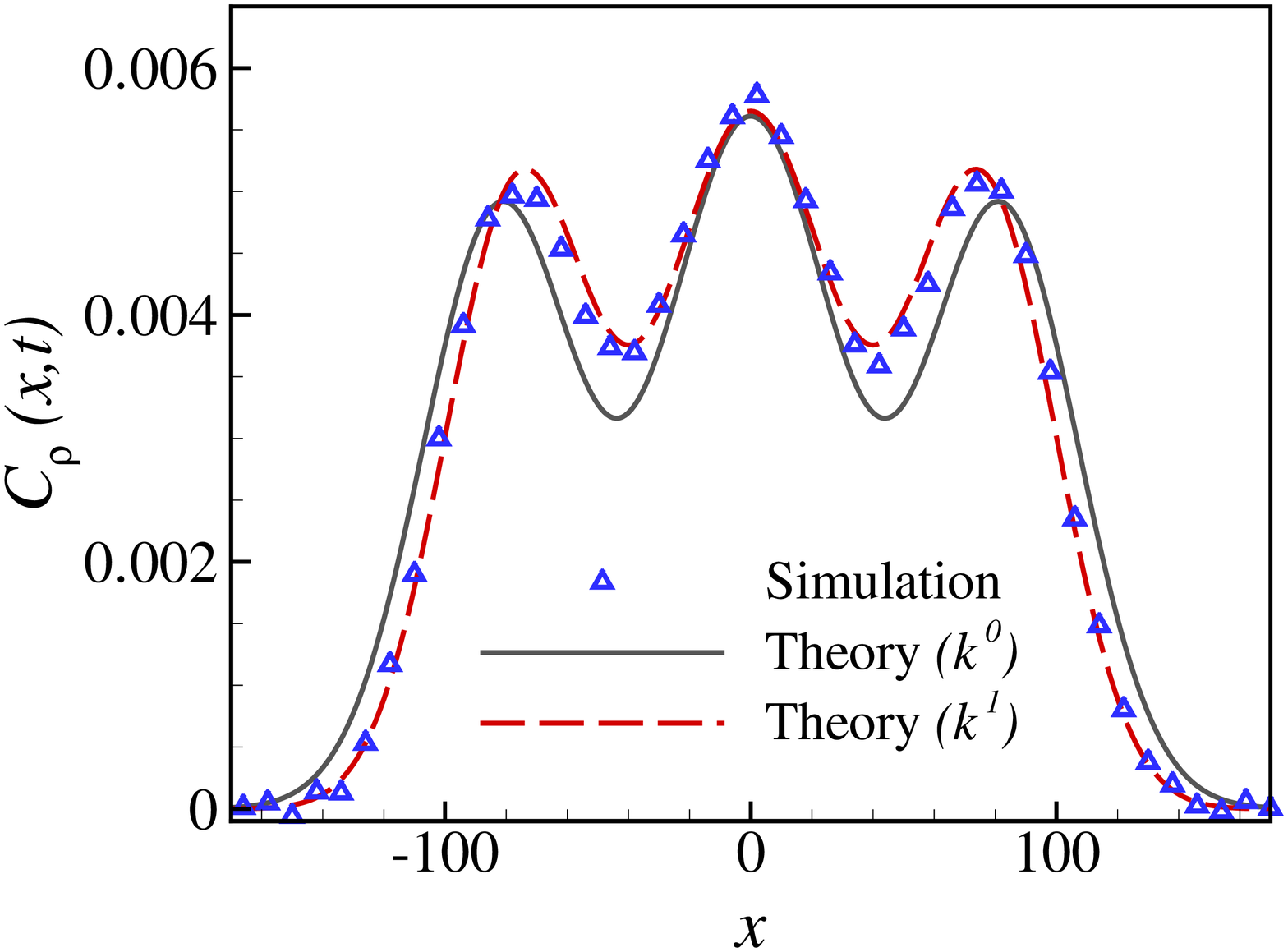} (d)
	\end{center}\end{minipage} \hskip0.01cm	
	\begin{minipage}[h]{0.32\textwidth}\begin{center}
		\includegraphics[width=\textwidth]{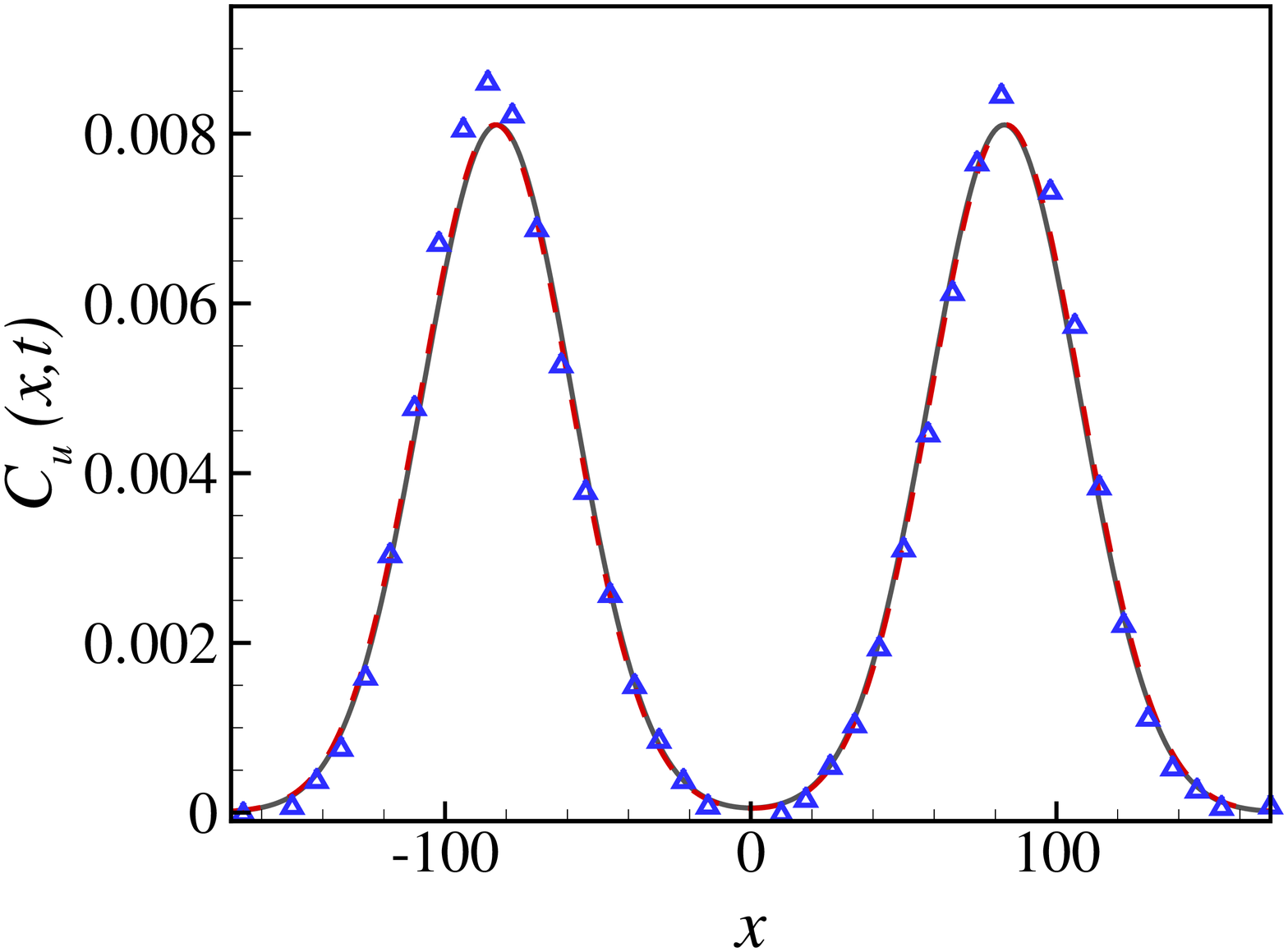} (e)
	\end{center}\end{minipage} \hskip0.01cm	
    \begin{minipage}[h]{0.32\textwidth}\begin{center}
		\includegraphics[width=\textwidth]{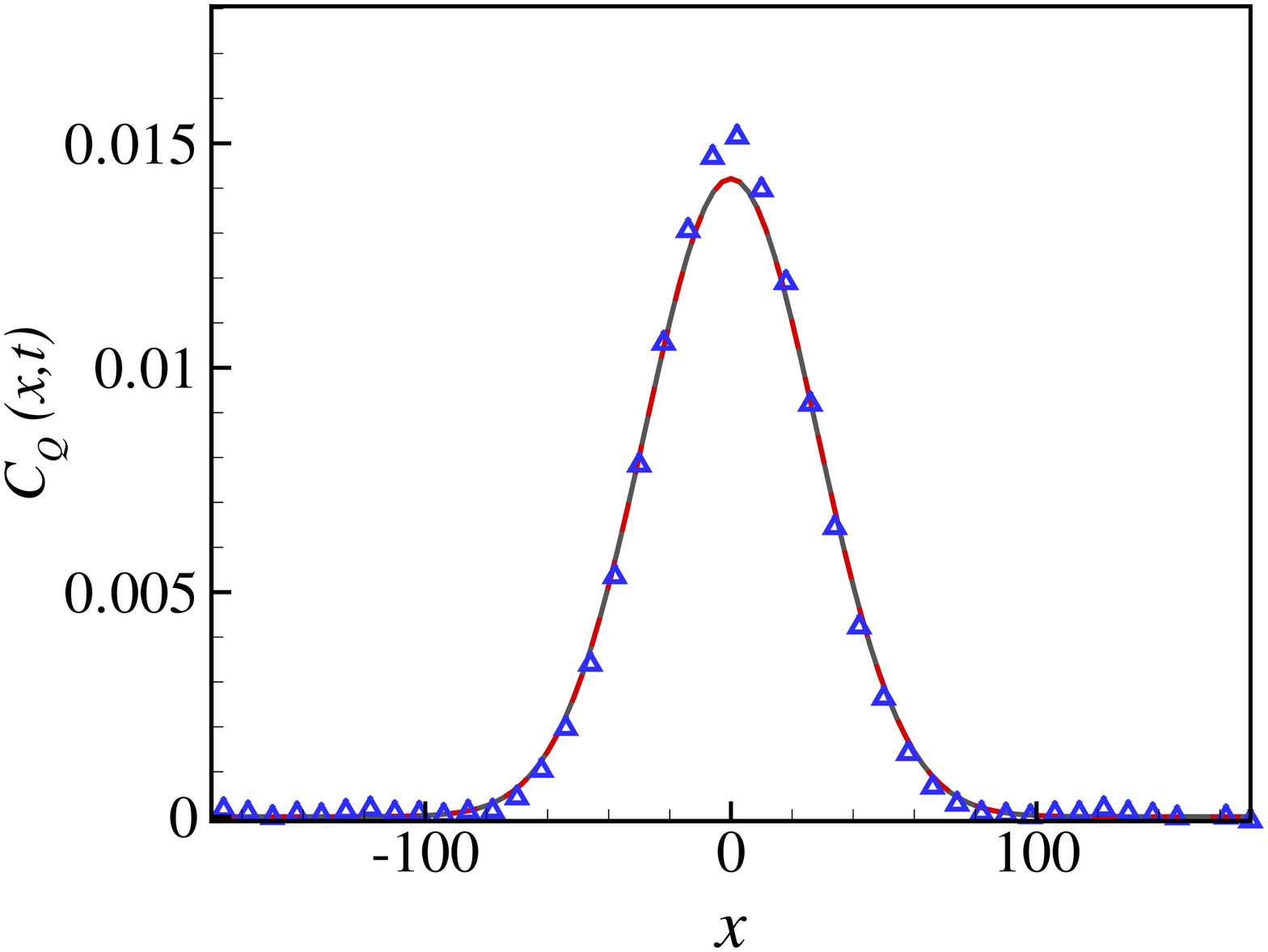}  (f)
	\end{center}\end{minipage} 	
\caption{(color online) Analytical correlation functions up to zeroth order in terms of amplitudes (solid black lines) for (a) density, (b) velocity and (c) heat fluctuations are compared to numerical results (symbols) for a system with $\rho=0.04$, $T=0.01$ and $MFT=33.8$ at $t=600, 1400$ and $2800$ in rescaled time units. (d)-(f) show first-order correlation profiles (dashed red lines) compared to their zeroth-order counterparts (solid gray lines) at $t=600$. Note that the correlation functions in all theories coincide in low-temperature limit and we have chosen to plot TKO's theory as a representative.}
\label{fig:low_temp_propag}
\end{center}\end{figure*}

Fig.\ref{fig:low_temp_propag}, shows density (a) momentum density (b) and heat correlation (c) profiles for a system of average density $\rho=0.04$, temperature $T=0.01$ and mean free time $MFT=33.8$. The measurements have been made at $t=600$ (blue triangles), $t=1400$ (red squares) and $t=2800$ (green circles) in rescaled time units to show the temporal evolution of the correlation functions in addition to their spatial dependence. The analytical correlation functions (solid lines) are given by Eqs. (\ref{e:corr_rho_BFT}-\ref{e:corr_q_BFT}) in which the amplitudes up to `zeroth order' with respect to the wave-number ($k$) are taken into account. For the sake of comparison, we have also plotted the `first-order' correction to the correlation functions (dashed lines) for $t=600$ that will be discussed below. Note that in the low temperature regime the hydrodynamic equations as well as the consequent correlation profiles reduce to the their Newtonian counterpart in all formalisms.

The generic behavior observed in ((a)-(c)) is that the agreement between theory and simulation improves as one goes further into hydrodynamic regime by increasing the observation times from $t=600$ to $t=2800$. For observation times above $t=2000$ (or $t\approx60~MFT$) our system indicates a good agreement between analytical results and numerical data (as expected) and thus confirms the accuracy of our model and simulation methods in the low density regime. At smaller time scales (e.g $t=600$), some deviations from analytical results are observed. The reason goes back to the contribution of high frequency short wavelength modes at these time scales, whose effects are not taken into account in Eqs.(\ref{e:corr_rho}-\ref{e:corr_Q}) where we have only kept the first lowest order terms. It is possible to account for the contribution of such short-wavelength effects by adding successive corrections to the zeroth-order correlation functions.

Fig.~\ref{fig:low_temp_propag}(d)-(f) compare the zeroth-order correlation profiles (solid gray lines) with the corresponding first-order expressions (dashed red lines) at $t=600$. The latter is obtained by keeping the first leading-order term in Eqs.~(\ref{e:corr_rho}-\ref{e:corr_Q}) that has been discussed in more detail in the Appendix. As Fig.~\ref{fig:low_temp_propag}(d) clearly shows, first-order correction to density profile slightly pushes the Brillouin peaks towards the center and thus improves the agreement between theory and simulation data. Nevertheless, such correction has negligible effect on velocity profile (panel(e)) and (as expected) does not change the heat profile. It appears that the role of short-wavelength corrections is more important in the propagation of density fluctuations, where thermal and sound modes are coupled together, than velocity and/or heat fluctuations in which information purely propagates through acoustic or thermal modes. Besides this subtle issue the difference between zeroth- and first-order correlation functions diminishes in all profiles by increasing time scales and reaching the hydrodynamic regimes (not shown here to preserve the clarity in figures).

Having checked the accuracy of our numerical simulations in different time scales in low temperature limit we now turn to the more interesting regime of intermediate and high temperatures.

\subsubsection{Intermediate and extremely relativistic regimes}
\begin{figure}[t!]\begin{center}
    \centering
	\includegraphics[width=7.5cm]{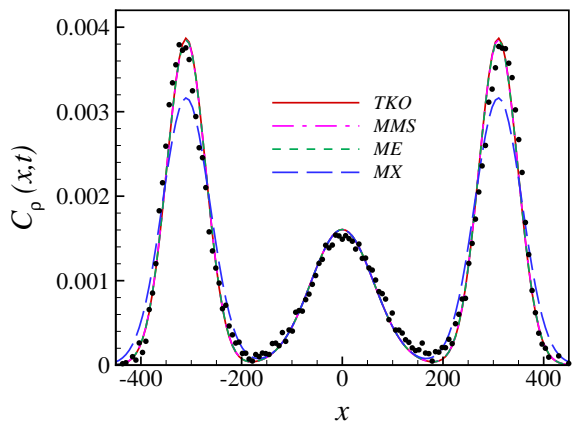}
\caption{(color online) Analytical correlation profile of density fluctuations obtained from TKO, MMS, ME and MX theories (colored lines) are compared to simulation results (black dots) for a system with $\rho=0.04$, $T=3$, $MFT=5.79$ and $t=500$. Note that the difference between ME, MMS, TKO theories is almost negligible at $T=3$ and cannot be observed in this figure.}
\label{fig:high_temp_propag}
\end{center}\end{figure}

\begin{figure}[t!]
\begin{center}
    \centering
	\includegraphics[width=7.5cm]{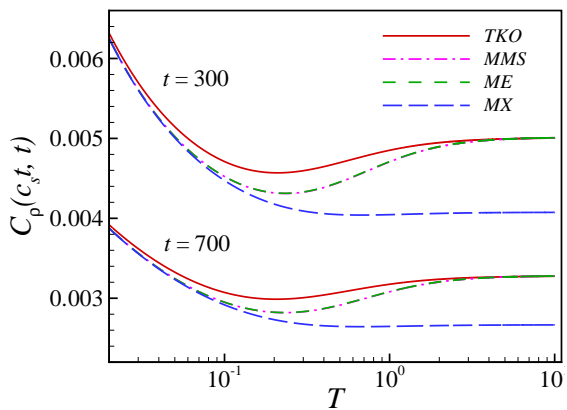}\\
\caption{(color online) Height of Brillouin peak in density correlation profile is plotted as a function of temperature for different formalisms for a system with $\rho=0.04$, at two different times.}
\label{fig:comp_height}
\end{center}
\end{figure}

\begin{figure}[t!]
\begin{center}
    \centering
	\includegraphics[width=7.5cm]{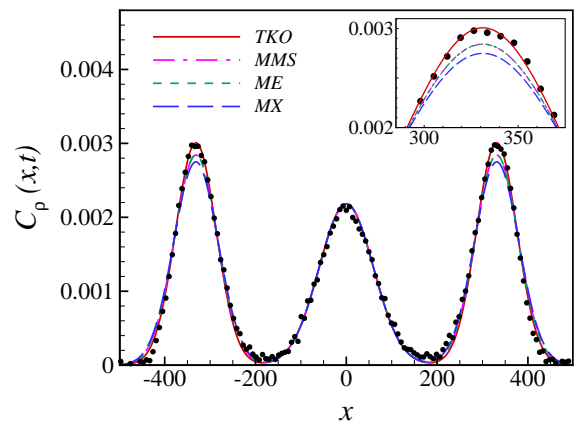}	
 \caption{(color online) Analytical density correlation profiles obtained from TKO, MMS, ME and MX theories (colored lines) are compared to simulation results (black dots) for a system with $\rho=0.04$, $T=0.2$ and $\text{MFT}=8.99$ which is observed at $t=700$ in rescaled time units. Inset shows details.}
\label{fig:int_temp}
\end{center}
\end{figure}

In this section, we only discuss results for density fluctuations as it suffices to discriminate between various hydrodynamic theories considered in this work. To this end we calculate Eq. \eqref{e:corr_rho_BFT} for various formalisms and compare them to the corresponding numerical density correlation spectrum obtained via Eq.~\eqref{e:correlation}. Fig.~\ref{fig:high_temp_propag} shows the result for a system of $\rho=0.04$ and $T=3$. The chosen observation time ($t\approx90~MFT$) is large enough to ensure that hydrodynamic assumptions are applicable and thus the zeroth order correlation functions (that we are using from here on ) are able to give a good description of the system. It is observed that ME, MMS and TKO theories provide good fit to numerical data while the prediction of MX formalism shows a significant deviation from the simulation data in the Brillouin peaks which characterize the acoustic (or sound) modes. This result which is also observed in velocity correlations spectrum (not shown here) would certainly disqualify the latter theory as an acceptable description of relativistic fluids; however, other theories are essentially identical for such parameter regime ($T=3$), and thus need to be studied in a wider range of thermodynamic parameters.

One may seek to find the condition of largest deviation between various theories focusing on the Brillouin peaks, which is expected to behave differently in various formalisms through the sound attenuation parameter, $\Gamma$, as discussed in section \ref{s:fluc_propag}.  Fig.~\ref{fig:comp_height} depicts the height of the right-moving peak ($C_\rho (x=c_s t,t)$) as a function of temperature for a system of $\rho=0.04$ at two different observation times, $t=300$ and $t=700$ in rescaled time units. One notes that different formalisms agree very well in low-temperature Newtonian limit (as expected) while they behave differently as temperature increases. In MX theory a monotonic decrease at intermediate temperatures ($0<T<1$) is followed by a saturating behavior at extremely relativistic regimes ($T\gg 1$), in contrast to TKO, MMS and ME theories which display a shallow dip at intermediate temperatures and interestingly converge to a single value at high temperature regime. The largest deviation between the latter theories is observed at intermediate temperatures where the equilibrium velocity distribution of a relativistic gas undergoes a morphological phase transition \cite{Men12,mon14}. We note that, this interesting coincidence between the critical temperature of a relativistic gas ($T_C=0.2$) and the behavior of Brillouin peak in intermediate temperatures calls for a deeper investigation which is out of the scope of this work.

In the light of the above results, one would expect that numerical data at intermediate temperatures (specially at the critical temperature, $T_C=0.2$) is able to differentiate between various theories and thus provides conclusive evidence to prove (or disprove) the validity of the underlying relativistic hydrodynamic equations presented in the context of first-order theories. The density correlation spectrum given in Fig.~\ref{fig:int_temp} for $T=0.2$, confirms that TKO theory gives the best description of our MD simulations. This result, which has also been confirmed for $T=1$ (not shown), provides a reasonable evidence in favor of TKO formalism due to the fact that our model is a fully relativistic one without adjustable parameters and/or probabilistic factors and thus should in principle yield numerically exact results.

Before closing our analysis some notes are in order. First, the height and width of Rayleigh peak which describe the thermal mode in propagation of density fluctuations are in good agreement with analytical predictions in all formalisms (see middle part of Fig.~\ref{fig:high_temp_propag} and Fig.~\ref{fig:int_temp}). This result is repeated in heat correlation profile given by Eq.~\eqref{e:corr_Q} where propagation of fluctuations are purely diffusive (not shown). It can therefore be concluded that despite their different definitions of heat flux, $q^\mu$, all formalisms give good description of the system as far as thermal mode is concerned.

Second, despite the fundamental differences in their postulates, ME and MMS formalisms have led to almost similar curves in the entire temperature regimes. This is because the sound attenuation parameter of the two theories (see Table 1), are nearly equal due to the very small bulk viscosity.

Third, the deviation between numerical results and analytical curves, observed in the position of Brillouin peaks at highly relativistic regimes (Fig.~\ref{fig:high_temp_propag}), should be traced back to the the errors that have been introduced to the theory via the sound velocity $c_s=\sqrt{\gamma/(\rho h \kappa_{\small T})}$ due to the approximative equation of state, heat capacity and isothermal compressibility that are adopted for hard-sphere gas (see Table 1 caption). This effect disappears in the low energy limit where sound velocity is not large enough to make a notable deviation (as is seen in the inset of Fig.~\ref{fig:int_temp}.) Therefore, it seems that the inconsistency in the position of peaks observed in high temperature regimes is not a fundamental issue and can be improved by making a more accurate estimation of $c_s$ in the analytical calculations.

Forth, one might wonder whether the relaxation-time corrections of dissipative process as are discussed in the context of second order theories would affect our results. It is known that such corrections would generally change the dispersion relation of the medium \cite{Rom10} and thus the speed and height of Brillouin peaks, which are our decisive factors to differentiate various theories. In order to check this, we have made an estimation of the uncertainty caused by relaxation time corrections in Appendix \ref{app_2} and observed that the effect does not change the results reported in Fig.\ref{fig:int_temp}, and thus the hydrodynamic limit behavior  which is the subject of present results. Nevertheless, one might observe notable departure from first-order theories in short time scales (comparable to such relaxation times), that is out of the scope of this work and will be discussed elsewhere \cite{Gho16}.

\section{Concluding remarks}
Over the past few decades various first-order relativistic hydrodynamic theories for non-perfect fluids have been proposed and studied extensively. Since such theories use distinctly different set of assumptions, they naturally lead to distinctly different predictions. A reliable means to test the prediction of such theories therefore seems of utmost importance. In this work we have provided a definitive step in this regard by studying the propagation of perturbations in various theories and comparing them with results from extensive numerical simulations of a fully relativistic molecular dynamics model of a hard-sphere fluid in a wide range of temperature regimes at low densities. Interestingly, we find that it is in the intermediate temperature regime where the predictions of such theories deviate most from each other. Our main result is that the recent theory of TKO is more in line with our numerical simulations than the previously studied theories such as MX, ME and MMS. However, we note that, our work has only considered various first-order theories in a static background in hydrodynamic limit. We must note that the consistency (or lack) of the first-order theories with our numerical results cannot automatically be extended to higher-order versions of such theories, as care must be taken in such generalizations.

Finally, our approach could be used to study such systems in higher densities than studied here, where we would expect that deviation between TKO and conventional theories become more pronounced. Furthermore, our approach could also be used to test various second-order theories along the lines of the present work. Another possible avenue is to check the accuracy of constitutive equations, e.g. shear or heat flow, by producing velocity or temperature gradients as in Ref.~\cite{Gho13}. Such possibilities are currently under investigation.

\section*{Acknowledgements}
A. Montakhab acknowledges the support of Shiraz University Research Council. M. Ghodrat acknowledges financial support from the Research Institute for Astronomy and Astrophysics of Maragha (RIAAM) under research No. 1/4165-117.
Results presented in  Appendix B were prepared in response to suggestions by
an anonymous referee whose contribution is acknowledged.

\appendix
\section{First-order correction to correlation functions}\label{app_1}

The dynamical equations (\ref{e:corr_rho}-\ref{e:corr_Q}) give a very good description of the system in hydrodynamic limit where high frequency long wavelength modes are considered. In order to account for the contribution of  short-wavelength effects that arise in short time scales one could add successive corrections to the aforementioned equations and thus obtain higher order correlation functions. For example, considering the first leading-order term in Eq. \eqref{e:corr_rho}; that is $k D_{\rho} \exp(-\Gamma k^{2} t)\sin(c_s k t)$,  one would get the following first-order correction to density correlation function in all formalisms:

\begin{eqnarray}\label{e:1stcorr}
\nonumber &&C_\rho^{(1)}=\\
\nonumber &&\frac{D_{\rho}(\Gamma,\chi)}{8\sqrt{\pi} (\Gamma t)^{3/2}} \left((x+c_{s} t)e^{-\frac{(x+c_{s} t)^{2}}{4\Gamma t}}-(x-c_s t)e^{-\frac{(x-c_{s} t)^{2}}{4\Gamma t}}\right),\\
\end{eqnarray}
in which $D_{\rho}=(\chi (\gamma -1 )+\Gamma)/\gamma c_{s}$ with $\chi$ and $\Gamma$ given in table \ref{table_1}.

The first-order correction to four-velocity correlation, $C_u^{(1)}$, has the same functional form as Eq.~\eqref{e:1stcorr} with different coefficient, $D_u(\Gamma,\chi)$, that goes to the classical expression $D_u=(\chi(\gamma-1)-\Gamma_{cl.})/c_s$ with $\Gamma_{cl.}=\Gamma_{MX}$, in low temperature limit. The full expressions of $C_u^{(1)}$ in various formalisms are straightforward to obtain following the standard methods used in Sec.~\ref{s:fluc_propag}. Also, the first leading term in the propagation of heat fluctuations (Eq.~\eqref{e:corr_Q}) is $\mathcal O(k^2)$ and therefore the resulting heat correlation function is not changed to first order.

\section{Relaxation time effects}\label{app_2}
\begin{figure}[t!]
\begin{center}
    \centering
	\includegraphics[width=7.5cm]{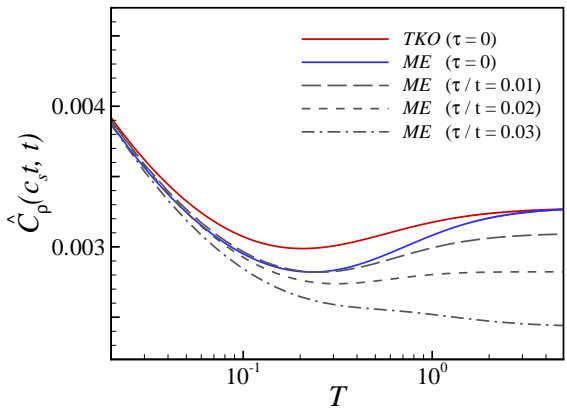}	
 \caption{(color online) The height of Brillouin peak in density correlation profile of first-order TKO and ME formalisms are compared to second-order ME formalism for two different ratios $\tau/t = 0.01, 0.02, 0.03$, in a system with $\rho=0.04$ and $t=700$. Note that the correction due to finite relaxation time ($\tau\neq0$) does not alter ME ($\tau=0$) results in such a way to make it comparable with TKO results.}
\label{fig:rlx_comp}
\end{center}
\end{figure}

The so called ``parabolic view" of the first-order theories is a reasonable approximation for dissipative processes that occur on hydrodynamic time scales, $t_{hyd.}\gg \tau_{rlx.}$. Nevertheless, if one considers relaxation time corrections in the context of second-order theories the acoustic modes are changed while the thermal mode remains unaffected. Since, the height of the former modes is a decisive factor in our arguments, we calculate the leading-order term due to relaxation time corrections in modified Eckart formalism and check whether it could improve the analytical correlation functions in this formalism.

In the simplest second-order theory of Israel-Stewart in Maxwell-Cattaneo form \cite{Rez13,Rom10}, the dispersion relation would read as:

\begin{equation}\label{e:dis_rel}
\omega=\pm i c_s k- \Gamma k^2 \mp i f_\tau k^3 +\mathcal{O}(k^4),
\end{equation}
with $\Gamma$ being the sound attenuation parameter as given in table \ref{table_1} and $f_\tau$ a real function of shear/bulk viscosity, thermal conductivity and the corresponding relaxation times due to these dissipative processes in the system (i.e., $\tau_\pi$, $\tau_\Pi$ and $\tau_q$). In this case the modified density distribution in Fourier space (corresponding to Eq.\ref{e:corr_rho}) is obtained as follows:

\begin{equation}\label{e:mod_corr_rho}
\frac{\hat{\rho}_k(t)}{\hat{\rho}_k(0)}=\frac{\gamma-1}{\gamma}e^{-\chi k^2 t} +\frac{1}{\gamma}e^{-\Gamma k^2 t}\cos(t k (c_s - f_\tau k^2)),
\end{equation}
whose back fourier transform gives the spatiotemporal evolution of density correlation function. Since the exact analytical expression was not available, we use the long wavelength approximation to replace the cosine term with the following expression:
\begin{eqnarray}\label{e:approx}
\nonumber\cos(&&t k(c_s- f_\tau k^2))\approx \\
&&(1 - \frac{1}{2} f_\tau^2 t^2 k^6)\cos(t k c_s)+f_\tau t k^3 \sin(t k c_s),
\end{eqnarray}
and obtain the leading order corrections to our first-order theory. In order to simplify the expression we have evaluated it at ($x=c_s t$) and arrive at
\begin{eqnarray}
\nonumber \hat{C}_{\rho}(&&x=c_s t,t)=\frac{1}{4\gamma \sqrt{\pi}\sqrt{\Gamma t}}\left(1-\frac{15}{16}\frac{f_\tau^2}{\Gamma^3 t}\right)\times\\
&&\left(e^{-\frac{c_s^2 t}{\Gamma}}+1\right)+e^{-\frac{c_s^2 t}{\Gamma}}\mathcal{F}(f_\tau,t,\Gamma,c_s),
\end{eqnarray}
with $\mathcal{F}(f_\tau,t,\Gamma,c_s)$ being a polynomial function of its arguments. As is seen, the above equation gives the height of right-moving Brillouin peak in first-order theories (Eq.(18)) and the leading order correction due to the relaxation times,
\begin{equation}\label{e:corr}
\delta \hat{C}_{\rho}^{(1)}(x=c_s t,t)=-\frac{1}{4\gamma \sqrt{\pi}\sqrt{\Gamma t}}\left(\frac{15}{16}\frac{f_\tau^2}{\Gamma^3 t}\right).
\end{equation}
All other terms exponentially decay to zero in the hydrodynamic limit, $t\rightarrow \infty$. As a first estimation of Eq.~\eqref{e:corr}, we assumed that all relaxation times are equal ($\tau_\pi=\tau_\Pi=\tau_q=\tau$) and the ratio $\tau/t=0.01, 0.02$ and $0.03$, in a system with $\rho=0.04$ and $t=700$. These are reasonable values according to the the lower limit of shear viscosity ($ \tau_\pi \geq 4\eta /(3\rho h(1-c_s^2))\approx0.36$) and the estimation given for massless hard sphere ($\tau_\pi=5/(3 n_0\sigma)\approx 52.7$) in \cite{Den11}. Figure \ref{fig:rlx_comp} indicates the results for the height of right-moving Brillouin peak as a function of temperature. As is seen, the inclusion of relaxation times in the context of Israel-Stewart second-order theory decreases the value of density correlation at $x=c_s t$, compared to the first-order ME formalism (except for very small values of $\tau/t$ where it has negligible positive effect in low temperature regime). Such a decrease in height could not improve the analytical results in modified Eckart formalism and therefore the results reported in Fig.~\ref{fig:int_temp} remains unchanged. Since the velocity of Brillouin peak would also increase as a result of such corrections (particularly in the high temperature regimes), the second order formalism might be able to improve the disagreement observed in Fig.~\ref{fig:high_temp_propag} between the position of Brillouin peaks in analytical and simulation results for some values of $\tau/t$.

Of course, a reliable argument of this kind would entail an exact solution of hydrodynamic equation in the context of second-order theories as well as a more accurate calculation of relaxation times for hard-sphere gas, which would be out of the scope of the present work. Nevertheless, the present argument suggest that the results reported here would not change by taking the relaxation times into account. One should, however, note that our results as well as the above argument are given in the hydrodynamic limit and should not be generalized to other conditions such as short time scales ($t\approx\tau_{rlx.})$ in which hyperbolic second order theories and relaxation time corrections are expected to become important.

\bibliographystyle{apsrev}
\bibliography{xbib}

\end{document}